\input{epsf}
\documentstyle[aps,preprint]{revtex}
\tightenlines
\begin{document}

\def\mbox#1#2{\vcenter{\hrule \hbox{\vrule height#2in
                \kern#1in \vrule} \hrule}}  
\def\sq{\,\raise.5pt\hbox{$\mbox{.1}{.1}$}\,}
\def\sqb{\,\raise.5pt\hbox{$\overline{\mbox{.1}{.1}}$}\,}
\def\simlt{\stackrel{<}{{}_\sim}}
\def\simgt{\stackrel{>}{{}_\sim}}

\title{Fractal Geometry of Quantum Spacetime at Large Scales}
\vspace{1.5 true cm}
\author{Ignatios Antoniadis \\
Centre de Physique Th\'eorique, Ecole Polytechnique,
\\F-91128 Palaiseau, France\\
{\it antoniad@cpht.polytechnique.fr}\\
\vspace{0.5 true cm}
Pawel O. Mazur\\
Dept. of Physics and Astronomy, University of S. Carolina,\\
Columbia, SC 29208 USA\\
{\it mazur@psc.psc.sc.edu}\\ 
\vspace{0.1 true cm}
and\\
\vspace{0.1 true cm}
Emil Mottola \\
Theoretical Division, Los Alamos National Laboratory, \\
MS B285, Los Alamos, New Mexico 87545 USA\\
{\it emil@pion.lanl.gov}}
\preprint{\begin{tabular}{r}  LA-UR-98-3410 \\ CPHT-S638.0898 \end{tabular}}
\maketitle

\begin{abstract}
\baselineskip 1 pc
We compute the intrinsic Hausdorff dimension of spacetime at the infrared
fixed point of the quantum conformal factor in 4D gravity.  The fractal
dimension is defined by the appropriate covariant diffusion equation in
four dimensions and is determined by the coefficient of the Gauss-Bonnet
term in the trace anomaly to be generally greater than $4$.  In addition to
being testable in simplicial simulations, this scaling behavior suggests a
physical mechanism for the screening of the effective cosmological
`constant' and inverse Newtonian coupling at very large distance scales,
which has implications for the dark matter content and large scale
structure of the universe.
\thispagestyle{empty}
\end{abstract}
\voffset=1.0 true cm
\newpage

\voffset=-0.2 true cm
\pagestyle{plain}
\pagenumbering{arabic}

The concept of gauge invariance, and indeed the term itself first entered
physics through the attempts of Weyl to extend the general coordinate
invariance of relativity to scale or conformal symmetry. Although these
early explorations led to models that were quickly ruled out by
observations, scale invariance has played a central role in several later
developments, most notably the impressive success of renormalization group
techniques in determining the critical behavior of second order phase
transitions in a wide variety of statistical systems. More recently,
conformal symmetry was rejoined with gravity in the quantum Liouville
theory of 2D surfaces where it has made possible the analytic solution of
the model \cite{KPZ}.

The key observation which made possible this progress in 2D gravity is that
the `classical' action is not complete, but rather must be augmented by an
additional term to account for the quantum trace anomaly of matter in a
general curved background. Since this additional term is infrared relevant
(strictly speaking, marginal) in the language of the renormalization group,
it cannot be neglected at large distances, where indeed it changes
completely the predictions of the classical theory and produces nontrivial
scaling exponents which are determined by the trace anomaly central charge
\cite{KPZ}. These analytic predictions of scaling exponents have been
confirmed by numerical simulations on dynamically triangulated lattices
\cite{DT}.

In four spacetime dimensions the quantum trace anomaly of massless
conformal fields requires the introduction of an additional term in the
effective action by precisely analogous reasoning \cite{Rie,am}. This term
is {\it nonlocal} when written in a generally covariant form, becoming
local only in the conformal gauge parameterization. Let us be very clear
that because of the existence of {\it non}trace modes in the four
dimensional metric tensor, this anomaly generated term is only one of many
in the full effective action and therefore gives only strictly limited
information about the quantum effects of matter in curved
spacetime. Certainly it cannot be used to extract any information about
Ward identities satisfied by products of {\it un}traced energy-momentum
tensors, or any other quantity which depends on the nontrace ({\it i.e.}
ordinary graviton) dynamics of the metric. Nor can this effective action be
trusted to describe the physics of the ultrashort Planck scale, where the
very notion of the spacetime metric itself most likely becomes
useless. Nevertheless, this term in the effective action reproduces the
{\it full} dependence on the trace of the metric {\it required} by the
trace anomaly of massless degrees of freedom, and just as in the 2D
Liouville theory it is strictly infrared marginal under conformal
transformations. Hence, it is sharply distinguished from any of the myriad
other terms in the quantum effective action of gravity by the fact that it
does {\it not} decouple at low energies and hence its effects cannot be
neglected on large distance scales.

The new term in the effective action of 4D gravity produces an important
qualitative departure from the classical Einstein theory, namely, the
addition of a {\it new} degree of freedom in the trace or conformal sector
of the metric. As is well known, the Einstein theory describes only spin-2
propagating degrees of freedom, its spin-0 or trace component being
completely frozen by the classical constraints \cite{York}. Since these
classical constraints of general relativity are necessarily modified by the
trace anomaly, the new fourth order term in the quantum effective action
allows the scalar spin-0 component of the metric tensor to become dynamical
and fluctuate as well. However, even in the quantum theory the constraints
of diffeomorphism invariance are quite stringent and serve to eliminate all
the purely local propagating degrees of freedom in the scalar sector
(negative metric or otherwise), leaving behind only a single new {\it
global} degree of freedom \cite{amm95}. This is in marked contrast to local
fourth order derivative modifications of the Einstein action which describe
local propagating modes that lead to violations of energy positivity and
unitarity.

Because the fluctuations of the conformal factor are global in character
the classical Einstein theory should remain largely intact at all scales
intermediate between the extreme ultraviolet Planck scale and the extreme
infrared horizon scale\footnote{An exception to this may be in very strong
gravitational fields, such as in black holes.}. In particular, there is no
new local gravitationally coupled spin-0 degree of freedom predicted by the
new term, and therefore no conflict with observational bounds on such
dilaton-like scalar particles. However we have argued elsewhere that the
fluctuations of the conformal factor become relevant at the horizon scale
of the classical expanding universe \cite{am,amm97}.  In fact, the
effective theory of the trace anomaly induced action predicts a conformally
invariant fixed point of gravity at very large distance scales where the
new term dominates the classical Einstein term, scale invariance is
restored, and the anomalous dimensions of the Einstein and cosmological
terms can be computed in closed form in terms of a single anomaly
coefficient $Q^2$, just as in the 2D Liouville theory.  In a sense this
restoration of global conformal symmetry at the infrared fixed point is a
realization of Weyl's original idea of `gauge' invariance in gravity, where
the rigid distance standard of Newtonian physics, which is preserved in
Einstein's theory, gives way to complete invariance under distance
reparameterizations, in the same way as in the theory of second order phase
transitions at the critical temperature.

In previous letters we have discussed the predictions of the anomalous
dimensions and scaling behavior of the infrared fixed point of the
conformal factor, both for numerical simulations on dynamically
triangulated lattices, and on the spectrum and statistics of the cosmic
microwave background radiation \cite{amm94,amm97}. In this Letter we wish
to expose two additional effects, namely the fractal dimension of spacetime
and the effective running of the Newtonian and cosmological `constants' at
large distances, both of which are implied by the anomalous dimensions and
scaling exponents computed previously. These provide additional predictions
for the simplicial simulations and suggest additional implications for the
dark matter content and very large scale structure of the universe, which
we discuss below.

\vspace{.5cm}
{\em Hausdorff Dimension}.\\ 
In order to define a Hausdorff dimension for
spacetime one needs to relate the geodesic distance $\ell(x,x')$ between
points $x$ and $x'$ to the volume $V_{\ell}$ enclosed by the spherical
surface with radius equal to $\ell$. The scaling relation between the two,
\begin{equation}
V_{\ell} \sim \ell^{d_H}
\label{hsdf}
\end{equation}
for large $\ell$ defines the {\it intrinsic} Hausdorff dimension $d_H$ of the
space.

In 2D quantum gravity the definition of distance that seems to be most
appropriate is that defined by the heat kernel of the Laplacian operator
$\sq$, {\it i.e.} \cite{Wat}
\begin{equation}
K_2(x, x'; s; g) = <x|\exp (-s\sq)|x'>\,.
\end{equation}
This quantity has the advantage of being manifestly covariant, and since
$\sq$ in the metric $g$ is an operator with well-defined scaling dimensions under conformal transformations,
\begin{equation}
\sq = e^{-2\sigma}\sqb \qquad {\rm for} \qquad g_{ab} = e^{2\sigma} \bar
g_{ab},
\label{ggbar}
\end{equation}
the heat kernel $K_2$ has a short distance expansion whose anomalous
scaling behavior may be calculated easily by standard techniques. We define
the average geodesic length squared that a scalar particle will have
diffused after proper time $s$ by
\begin{equation}
\ell^2_s \equiv {1\over V}\left\langle \int d^2x \sqrt g \int d^2x' \sqrt g'\,
\ell^2(x,x';g) K_2(x,x';s;g)\right\rangle_{_V}
\end{equation}
where the average is with respect to the fixed volume Liouville partition
function. By expanding the heat kernel $K_2$ in a power series in $s$ it is easy
to see that
\begin{equation}
\ell^2_s \sim s \qquad {\rm as} \qquad s \rightarrow 0
\label{dif}
\end{equation}
which is the standard result for a particle undergoing Brownian motion.

On the other hand the probability that the particle will come back to a
small region within $\epsilon$ of its starting point $x_0$ after proper
time $s$ is
\begin{equation}
\left\langle \int d^2x \sqrt g K_2(x,x_0;s;g) 
f_{\epsilon}(x,x_0)\right\rangle_{_V} =
\left\langle \int d^2x \sqrt g (1 - s \sq + {\cal O}(s^2))
f_{\epsilon}(x,x_0)\right\rangle_{_V}
\label{comeback}
\end{equation}
where $f_{\epsilon}(x,x')$ is any smooth function with support only for
$\ell(x,x') \sim |x-x'| \simlt \epsilon$ normalized by $\int d^2 x \sqrt g
f_{\epsilon}(x,x') = 1$, {\it i.e.} it approaches the delta function,
${1\over \sqrt g} \delta^2 (x-x')$ as $\epsilon \rightarrow 0$. From this
factor of ${1\over \sqrt g}$ and (\ref{ggbar}) it follows that the operator
multiplying the term linear in $s$ in (\ref{comeback}) is a density of
weight $-1$ in the Liouville theory,
\footnote{Here we follow the standard conventions in $D=2$ in which
the conformal weights are chiral and hence should be doubled to
compare with the corresponding weights in $D=4$.}
and therefore has a well-defined scaling behavior with volume. Hence,
\begin{equation}
s\left\langle \int d^2x \sqrt g \sq 
f_{\epsilon}(x,x_0)\right\rangle_{_V} =
s\left\langle \int d^2x \sqrt{\bar g} e^{2\alpha_{-1}\sigma}\sqb
\bar f_{\epsilon}(x,x_0)\right\rangle_{_V} 
\sim s V^{\alpha_{-1}\over \alpha_1}\,,
\label{fscal}
\end{equation}
where the finite volume scaling of the last proportionality follows by a
simple constant shift in the Liouville field, $\sigma \rightarrow
\sigma + \sigma_0$, and the $\alpha_n$ are scaling dimensions, 
given in the Liouville theory by the general formula,
\begin{equation}
\alpha_n = n + {\alpha_n^2\over Q^2} = {2n \over 1 + \sqrt{1 - {4n \over Q^2}}}
\label{alpt}
\end{equation}
in terms of $Q^2$ which is determined in terms of the matter central charge
(anomaly coefficient) $c_m$ by
\begin{equation}
Q^2 = {25 - c_m\over 6}\,,\qquad D=2\,.
\end{equation}

Since the return probability (\ref{comeback}) is independent of a rescaling
of the total volume $V\rightarrow \lambda V$ for small $s$ (before the
particle can feel the effects of the finite volume which is assumed to be
very large, {\it i.e.} lattice finite volume effects are assumed to be
irrelevant), it follows that $s$ must scale like
\begin{equation}
s \rightarrow \lambda^{-{\alpha_{-1}\over \alpha_1}} s
\end{equation}
under a global volume scaling. From this result and (\ref{dif}) we conclude
that
\begin{equation}
\ell^2_s \sim s \sim V_{\ell}^{-{\alpha_{-1}\over \alpha_1}}\,.
\end{equation}
Inverting this relation for $V_{\ell}$ in terms of $\ell$ and using
(\ref{hsdf}) gives finally the Hausdorff dimension,
\begin{equation}
d_H = -2\,{\alpha_1\over\alpha_{-1}} = 2 {\sqrt{25-c_m} + \sqrt{49-c_m}\over
\sqrt{25-c_m} + \sqrt{1-c_m}} \ge 2, \qquad D=2
\label{dHtwo}
\end{equation}
for 2D gravity \cite{Wat}. This formula predicts $d_H = 4$ for pure 2D
gravity ({\it i.e.} $c_m = 0$), $d_H = 2 (1 + \sqrt 2)$ at the limit of its
validity for $c_m =1$ and $d_H \rightarrow 2$, its classical value, in the
classical limit, $c_m \rightarrow -\infty$, or $Q^2 \rightarrow + \infty$,
indicating that the geometries become smooth and classical in that limit. A
summary of the agreement between (\ref{dHtwo}) and numerical simulations
for various values of $c_m$ is given in refs. \cite{Amb}.

A precisely parallel computation will now be given for four dimensional
gravity at the infrared fixed point determined by the fluctuations of the
quantum conformal factor. First we note that we cannot use the second order
operator $\sq$ in four dimensions since it no longer has a well-defined
conformal weight.  Adding a $-R/6$ correction does not help since it
operates on conformal scalars which have weight one in $D=4$ (not zero as
in $D=2$).  Thus the calculation above will not go through as it did in
$D=2$ for any second order scalar operator in $D=4$.  Instead we suggest
that the appropriate differential operator with which to define an
invariant diffusion length is
\begin{equation}
\Delta_4 \equiv{\sq}^2 + 2R^{ab} \nabla_a \nabla_b - 
\hbox{$2 \over 3$} R {\sq}+ \hbox{$1 \over 3$}(\nabla^a R)\nabla_a
\end{equation}
which satisfies
\begin{equation}
\Delta_4= e^{-4\sigma} \bar{\Delta}_4
\end{equation}
for $g_{ab} = e^{2\sigma}\bar g_{ab}$. Hence like $\sq$ in $D=2$,
$\Delta_4$ in $D=4$ transforms homogeneously under local conformal
rescalings of the metric and operates on scalar functions with zero
conformal weight. It is also precisely the operator which enters the
anomaly induced action in four dimensions, just as $\sq$ does in two
dimensions. Hence we propose to define the heat kernel,
\begin{equation}
K_4(x, x'; s; g) \equiv <x|\exp (-s\Delta_4)|x'>\,
\label{defK}
\end{equation}
and the average geodesic length after proper time $s$ by
\begin{equation}
\ell^4_s \equiv {1\over V}\left\langle \int d^4x \sqrt g \int d^4x' 
\sqrt g' \ell^4(x,x';g) K_4(x,x';s;g)\right\rangle_{_V}
\label{ellf}
\end{equation}

The angular brackets here denote the quantum expectation value in the
finite volume Euclidean partition function $Z(\kappa ; V)$, namely,
\begin{equation}
\left\langle {\cal O} \right\rangle_{_V} = {1\over Z(\kappa ;V)}
\int [{\cal D}\sigma]\,{\cal O}\,\exp \left( - S_4[\sigma]
-{1\over 2\kappa}S_2[\sigma]\right)
\delta\left( S_0[\sigma] - V\right)\,,
\end{equation}
where
\begin{equation}
S_0 [\sigma] = \int d^4 x\,\sqrt{g} = \int d^4 x\, 
\sqrt{\bar g}\,e^{4\sigma}\,.
\label{vol}
\end{equation}
is the classical Euclidean four-volume,
\begin{equation}
S_2 [\sigma] = -\int d^4 x\,\sqrt{g}\, R = 6\int d^4x\,
\sqrt{\overline g}\,e^{2\sigma} \left[\sqb\sigma + 
(\overline{\nabla}\sigma)^2 - {{\overline R}\over 6}\right]
\label{EH}
\end{equation}
is the classical Einstein-Hilbert action, and
\begin{equation}
S_4[\sigma] = {Q^2 \over (4 \pi)^2}\int d^4 x \sqrt{g} 
\bigl[\sigma  {\Delta}_4 \sigma + \hbox{$1\over 2$}
\bigl(G - \hbox{$2 \over 3$} \sq R \bigr) \sigma \bigr]\ ,
\label{act}
\end{equation}
is the new term in the effective action required by the trace anomaly.  We
recall from refs. \cite{am,amm94} that the classical expressions for $S_0$
and $S_2$ are gravitationally dressed so that their classical scaling
dimensions ($4$ and $2$ respectively) are modified at the quantum level. In
the expression for $S_4$, $Q^2$ is the coefficient of the Gauss-Bonnet term
$G$ in the four dimensional conformal anomaly, given by
\begin{equation}
Q^2 = {1 \over 180}(N_S + \hbox{$11\over 2$} N_{WF} + 62 N_V - 28) +
Q^2_{grav}\ ,
\label{cent}
\end{equation}
in terms of the number of free scalar ($N_S$), Weyl fermion ($N_{WF}$) and
vector ($N_V$) fields, while the $-28$ and $Q^2_{grav}$ are the
contributions of the spin-0 conformal factor and spin-2 graviton fields of
the metric itself.

Now by expanding the heat kernel $K_4$ in (\ref{defK}) and (\ref{ellf}) in
powers of $s$ we obtain
\begin{equation}
\ell^4_s \sim s \qquad {\rm as} \qquad s \rightarrow 0
\label{difour}
\end{equation}
in place of (\ref{dif}). Defining next the return probability analogous to
(\ref{comeback}) in $D=4$,
\begin{equation}
\left\langle \int d^4x \sqrt g K_4(x,x_0;s;g) 
f_{\epsilon}(x,x_0)\right\rangle_{_V} =
\left\langle \int d^4x \sqrt g (1 - s \Delta_4 + {\cal O}(s^2))
f_{\epsilon}(x,x_0)\right\rangle_{_V}
\label{comebackf}
\end{equation}
where $f_{\epsilon}$ approaches ${1\over \sqrt g} \delta^4(x,x')$ as
$\epsilon \rightarrow 0$, we find that we need the conformal scaling
exponents $\beta_0$ and $\beta_8$, corresponding to the volume operator (a
density of weight $4$, codimension $\bar \Delta =4-w=0$) and $\Delta_4$
operator (a density of weight $w=-4$, codimension $\bar\Delta = 4 -(-4)
=8$) respectively.  Indeed for any operator with well-defined scaling
codimension $\bar\Delta$ in the absence of gravitational fluctuations we
have found previously that \cite{amm94}
\begin{equation}
\left\langle {\cal O}_{\bar\Delta} \right\rangle_{_V} \sim 
V^{1 - {\Delta\over 4}} = V^{\beta_{\bar\Delta}\over \beta_0}\ .
\label{Oscal}
\end{equation}
These scaling exponents $\beta_{\bar\Delta}$ are given by the general 
quadratic relation,
\begin{equation}
\beta_{\bar\Delta} = 4 -\bar\Delta + {\beta_{\bar\Delta}^2\over 2Q^2}\,,
\label{betaf}
\end{equation}
analogous to (\ref{alpt}), in the notation of \cite{amm94}, where
$\bar\Delta = 4-w$ is the codimension of the operator with conformal weight
$w$ (analogous to $2-2n$ in $D=2$, so that $2\alpha_n$ is replaced by
$\beta_{\bar\Delta}$ in $D=4$).  Hence we find that the term linear in $s$
in (\ref{comebackf}) scales like
\begin{equation}
\left\langle \int\, d^4x\, \sqrt g\, s\, \Delta_4\,
f_{\epsilon}(x,x_0)\right\rangle_{_V} = \left\langle \int\, d^4x\, 
\sqrt {\overline g}\, s\, \overline\Delta_4 e^{\beta_8\sigma}
\overline f_{\epsilon}(x,x_0)\right\rangle_{_V} \sim s V^{\beta_8\over \beta_0}
\label{fscalf}
\end{equation}
so that $s$ must scale like
\begin{equation}
s \rightarrow \lambda^{-{\beta_8\over \beta_0}} s
\end{equation}
under a global volume scaling, $V\rightarrow \lambda V$ in $D=4$. Combining
this result with (\ref{difour}) we obtain
\begin{equation}
\ell^4_s \sim s \sim V_{\ell}^{-{\beta_8\over \beta_0}}\,,
\end{equation}
so that solving for $V_{\ell}$ in terms of $\ell$ and using (\ref{hsdf})
and (\ref{betaf}) yields the Hausdorff dimension,
\begin{equation}
d_H = -4 \ {\beta_0\over \beta_8} = 4\,\,{1 + \sqrt{1 + {8\over Q^2}}\over
1+ \sqrt{1 - {8\over Q^2}}} \ge 4, \qquad D=4\,.
\label{dhf}
\end{equation}
Thus the fractal dimension of four dimensional quantum spacetime is also
generally greater than its classical value.

The value of $Q^2$ is at present uncertain, principally because of the
unknown infrared contribution of gravitons $Q^2_{grav}$ in (\ref{cent}),
which is likely to be close to $8$ \cite{amm92}. If the total $Q^2 < 8$ we
have argued in a previous letter \cite{amm94} that the theory undergoes a
BKT-like transition to a phase dominated by a dense `gas' of long extruded
structures, similar to the branched polymer phase seen in both the $D=2$
(for $c_m >1$) and $D=4$ simplicial simulations \cite{DT4}.  In that case
the typical spacetimes become so irregular that apparently not even a
fractal Hausdorff dimension adequately describes them. This is reflected in
the square root in the denominator of (\ref{dhf}) becoming imaginary for
$Q^2 < 8$. At $Q^2 =8$, $d_H = 4 (1 + \sqrt 2)$ while
\begin{equation}
d_H \rightarrow 4 + {16\over Q^2} + {\cal O}\left(Q^{-4}\right)
\qquad {\rm as} \qquad Q^2 \rightarrow
\infty \,,
\end{equation}
Thus the fractal dimension approaches its classical value in the limit in
which the fluctuations of the conformal factor are suppressed.  The
behavior of the Hausdorff dimension as a function of $Q^2$ for $Q^2 \ge 8$
is shown in Fig. 1.

\vspace{.5cm}
{\em Screening of the Cosmological and Inverse Newtonian Couplings}.

The quantum fluctuations of the conformal factor which give rise to the
nontrivial Hausdorff dimension are also responsible for gravitational
`dressing' of the volume and Einstein terms in the effective action.  The
classical expressions for $S_0$ and $S_2$ are replaced by corresponding
operators ${\cal O}_{\bar \Delta}$ with well-defined scaling dimensions
$\beta_0$ and $\beta_2$ of (\ref{betaf}) for $\bar\Delta = 0$ and $2$
respectively.  The corresponding couplings $\lambda =\Lambda/8\pi G_N$ and
$\kappa^{-1} = (8\pi G_N)^{-1}$ must scale inversely with the volume in
order for these terms in the effective action to be strictly marginal
deformations at the conformal fixed point. This is a consequence of the
consistency condition any covariant theory of gravity must satisfy when the
quantum fluctuations of the conformal factor are considered.

Since $\Lambda/G_N$ multiplies the volume it must scale like
\begin{equation}
{\Lambda\over G_N} \sim V^{-1}\,.
\label{lf}
\end{equation}
Since $(8\pi G_N)^{-1}$ multiplies an operator with conformal weight
$\beta_2$ its finite volume scaling is
\begin{equation}
G_N \sim V^{\beta_2\over \beta_0} \equiv V^{\delta}
\label{Gscal}
\end{equation}
We note that in this case the operator which becomes $S_2$ in the classical
limit $Q^2 \rightarrow \infty$ also acquires an additive renormalization at
the fixed point, as discussed in refs. \cite{am,amm94}.

The finite volume scaling relations (\ref{lf}) and (\ref{Gscal}) are
predictions for 4D simplicial simulations, but are not yet directly
relevant for continuum physics since the units in which the volume $V$ is
measured have not been specified. Hence it is always possible to absorb
this scaling of dimensionful quantities into a constant shift in $\sigma$.
For a meaningful comparison a dimensionless ratio of two quantities having
zero naive engineering dimensions should be formed so that one is measured
in units of the other. Such a quantity is the dimensionless coupling,
\begin{equation}
G_N\Lambda \sim V^{2\delta -1} \sim \ell^{d_H (2\delta -1)}
\label{lamscal}
\end{equation}
upon incorporating our previous result for the Hausdorff dimension. This
relation measures the cosmological term in units of the Planck mass. Since
we have from (\ref{betaf}) and (\ref{Gscal}),
\begin{equation}
2\delta - 1 = {\sqrt{1 - {8\over Q^2}} - \sqrt{1 - {4\over Q^2}}\over
1 + \sqrt{1 - {4\over Q^2}}} < 0
\end{equation}
for $Q^2 > 8$, the effective cosmological `constant' in units of the Planck
mass {\it decreases} at large distances, and $G_N\Lambda
\rightarrow 0$
at the infrared fixed point in the infinite volume limit. Thus the quantum
fluctuations of the conformal factor in 4D gravity provides a mechanism for
the effective screening of the cosmological coupling at large distances,
independently of its value from microscopic physics.

One may also compare the running of the Newtonian and cosmological
couplings to a length scale fixed by a matter field operator, such as
$\int\, d^4x\,\sqrt g\, \bar\psi\psi$. Since the fermion field has
dimension $3\over 2$ this operator gets gravitationally dressed by the
exponent $\beta_3$. Converting the finite volume scaling of this operator
according to (\ref{Oscal}) to its corresponding coupling, we find that the
fermion mass $m$ scales with the volume like
\begin{equation}
m \sim V^{-{\beta_3\over \beta_0}}
\end{equation}
This provides an independent standard of length against which we can
measure the running of $G_N$ and $\Lambda$. Then we find
\begin{eqnarray}
G_Nm^2 &\sim &V^{\delta - 2{\beta_3\over \beta_0}} = V^{\beta_2 
-2\beta_3\over \beta_0}
\sim \ell^{d_H{\beta_2 -2\beta_3\over \beta_0}} \nonumber\\
{\Lambda \over G_Nm^4} &\sim &V^{-1 + 4{\beta_3\over \beta_0}} \sim
\ell^{d_H {4\beta_3-\beta_0\over \beta_0}}\,.
\label{Gscalm}
\end{eqnarray}
Since
\begin{equation}
d_H {\beta_2 -2\beta_3\over \beta_0} = 2 \left(1 + 
\sqrt{1 + {8\over Q^2}}\right)
\left[{1\over 1 + \sqrt{1 - {4\over Q^2}}} - {1 \over 1 +
\sqrt{1 - {2\over Q^2}}}\right] > 0
\label{Gscalp}
\end{equation}
for $Q^2 > 8$ the Newtonian `constant' {\it increases} at large distances.
This implies that the gravitational attractive force between two fermions
falls more slowly than the classical Newtonian $r^{-2}$ force.

On the other hand since
\begin{equation}
d_H {4 \beta_3 - \beta_0\over \beta_0} = 4
\left(1 + \sqrt{1 + {8\over Q^2}}\right)\left[{1\over
1 + \sqrt{1 - {2\over Q^2}}} - {1 \over 1 + \sqrt{1 - {8\over Q^2}}}\right] <0
\end{equation}
for $Q^2 > 8$ we conclude that the the cosmological `constant' measured in
units of the fermion mass decreases and is effectively screened at large
distances.

For large $Q^2$ the large distance power laws in all these cases simplify
and we have
\begin{eqnarray}
G_N\Lambda &\sim &\ell^{-{4\over Q^2}} \nonumber\\
G_Nm^2 &\sim & \ell^{+{1\over Q^2}} \nonumber\\
{\Lambda \over G_Nm^4} &\sim &\ell^{-{6\over Q^2}}, \qquad Q^2 \gg 8
\end{eqnarray}

We emphasize that all these scaling relations, and in particular their
dependence on the power $Q^2$ given by (\ref{cent}) have been derived using
the anomalous dimensions that apply at the infrared fixed point of the
conformal factor.  Because the fluctuations of $\sigma$ which give rise to
these anomalous dimensions are suppressed inside the scale of the horizon,
the screening effects described here should become evident only at the very
largest scales of the universe, comparable to the horizon scale, of order
$1000$ Mpc or greater.

\vspace{.5cm}
{\em Consequences for Cosmology}.\\ The effective screening of the
cosmological `constant' at large distances suggests that it may be possible
to construct a cosmological model in which the vacuum energy component runs
continuously to smaller values as the universe expands. Consider a FRW
model with expansion determined by a scale factor with an arbitrary power,
$\nu$.  Then the physical geodesic length $\ell$ of the previous treatment
should scale like
\begin{equation}
\ell \sim a(t) \sim t^{\nu}\,.
\label{frw}
\end{equation}
From (\ref{lamscal}) this implies that the vacuum energy density
$\rho_{\Lambda}$
scales like
\begin{equation}
\rho_{\Lambda} \equiv {\Lambda \over 8\pi G_N}\sim M_{pl}^4\ 
t^{\nu d_H (2\delta -1)}\,,
\end{equation}
when measured in Planck units (hereafter fixed). For any $\nu$ the Hubble
`constant' scales as
\begin{equation}
H(t) \equiv {\dot a \over a} \sim t^{-1}\,,
\end{equation}
so if Einstein's equations hold, or less stringently, if the Ricci scalar
remains proportional to the cosmological term in the effective equations of
motion, then we should have
\begin{equation}
R(t) \sim H^2 (t) \sim t^{-2} \sim G_N\rho_{\Lambda} \sim M_{pl}^2\ t^{\nu
d_H (2\delta -1)}\,.
\label{curv}
\end{equation}
Therefore the expansion power of the scale factor in the cosmological model
is determined by the previous considerations to be
\begin{equation}
\nu = {2 \over d_H (1 - 2 \delta)} = {1\over 2}\
{ 1 - \sqrt {1 + {8\over Q^2}} \over 1 +   \sqrt {1 - {8\over Q^2}} 
- 2  \sqrt {1 - {4\over Q^2}}} \ge {1\over 2}\,.
\label{apower}
\end{equation}
This power equals $1\over 2$ at $Q^2 =8$, and goes to infinity linearly as
$Q^2 \rightarrow \infty$. The latter is the classical limit in which the
fluctuations of the conformal factor are suppressed and we recover de
Sitter spacetime with its exponential expansion. The behavior of $\nu$ as a
function of $Q^2$ is illustrated in Fig. 2.

This power law model has some interesting consequences. First, it implies
that for any $Q^2$ the vacuum energy density will always be a finite
fraction of the critical density,
\begin{equation}
\Omega_{\Lambda} \sim {\rho_{\Lambda}(t)\over H^2(t)} = constant
\label{olam}
\end{equation}
which one would generically expect to be of order unity. Indeed, if we
write
\begin{equation}
\Lambda (t) = k M_{pl}^2 \left({t_{pl}\over t}\right)^2
\end{equation}
and estimate the present age of the universe at $t_0 \simeq 12$ Gyr, and
the present Hubble parameter $H_0 = 75$ km/sec/Mpc we obtain
\begin{equation}
\Lambda_0 \simeq 2.05\, k \times 10^{-122} \, M_{pl}^2
\end{equation}
and
\begin{equation}
\Omega_{\Lambda} =\Lambda_0/ 3H_0^2 \simeq  0.40\, k
\end{equation}
which for $k$ of order unity is of order of the present bounds on the
cosmological term.  If $k = 1.5$ then $\Omega_{\Lambda} = 0.6$ as suggested
by a number of recent papers in the astrophysics literature \cite{cos}.
This would make it possible to have an asymptotically spatially flat
cosmological model with $\Omega_{total} =1$.  We also remark that if $\nu >
{2\over 3}$ this would imply an older age for the universe than that
inferred from classical matter dominated cosmology.

The second interesting aspect of the power (\ref{apower}) is that it rises
very rapidly from its minimum value of $1\over 2$, reaching $2\over 3$ at
$Q^2 \simeq 8.111$ and is greater than unity if $Q^2> Q^2_1 \simeq 8.636$.
Since $Q^2$ depends on the number of massless fields through (\ref{cent})
and `massless' in this context means light compared to $H(t)$, the
effective value of $Q^2$ could very well have been much higher in the early
universe than it is in the present epoch. A value of $Q^2 > Q_1^2$ implies
power law inflation, which if it persists for a long enough time removes
the `horizon problem' of the classical FRW cosmology.  Since the universe
becomes arbitrarily flat at late times in a model with scaling behavior
given by (\ref{curv}) and (\ref{olam}) there is no `flatness problem' in
such a model. Of course, a full cosmological model incorporating the
scaling behavior conjectured here (and free of any new problems) must be
constructed to realize this dynamical solution of the cosmological
`constant' problem.

As the universe expands and ages each light but not strictly massless
degree of freedom will have its Compton wavelength become of order of the
horizon at a different time, when we must expect the possibility that the
`constant' in (\ref{olam}) might change, analogous to the changes in a
renormalization group beta function as kinematic mass thresholds are
crossed. The detailed physics of these transition eras would fix the value
of $\Omega_{\Lambda}$ during the next epoch of power law expansion.
Finally, from perturbative estimates and numerical simulations there is
reason to believe that the present value of $Q^2$ could be close to (but
presumably slightly larger than) $8$. If that is the case then the present
value of the power $\nu$ is somewhat larger than $1\over 2$, and perhaps
close to the $2\over 3$ predicted by standard matter dominated cosmology
(cf. Fig. 2).

Globally, of course, a cosmological model constructed along the lines of
the foregoing speculations would be quite different, and observationally
distinguishable from the standard model. First, the presence of the
effective cosmological term would be expected to modify the linear Hubble
relation at the very largest scales, which is a subject of very active
current interest. Secondly, because of the increase of the Newtonian
coupling at the largest scales according to (\ref{Gscalm}), estimates of
mass through virial arguments could be somewhat larger than the {\it
actual} mass present in large clusters and superclusters of
galaxies. Taking $Q^2 \simeq 8$ the power in (\ref{Gscalm}) and
(\ref{Gscalp}) is approximately $0.24$. Since $10^{0.24} \simeq 1.74$ the
increase of the Newtonian coupling over one decade of distance would lead
to a $74\%$ increase in the dark `matter' attributed to the supercluster by
virial estimates, on top of the effect of the homogenous vacuum density of
$\rho_{\Lambda}$.  Together, these two effects may eliminate the need to
invoke exotic forms of dark matter to close the universe, so this is a kind
of null prediction.  In addition to these effects, since the spectral index
of the initial adiabatic density fluctuations is also altered by the
infrared fixed point of the conformal factor \cite{amm97}, models of
structure formation would have to be reconsidered {\it ab initio}.
Finally, the most radical departure from the standard model arises from the
fractal structure of spacetime itself at the largest scales, which if
correct, implies that the universe and the matter distributed in it are not
homogeneous, even at scales approaching the horizon scale.

\vspace{1cm}
\noindent{\bf Acknowledgments}

The authors wish to acknowledge NATO grant Collaborative Research Grant
$971650$ for making this work possible. P. O. M. and E. M. also wish to
thank the Centre de Physique Th\'eorique at the Ecole Polytechnique for its
hospitality.

\vspace{1.5cm}

\newpage

\epsfxsize=10cm
\epsfysize=9cm
\centerline{\epsfbox{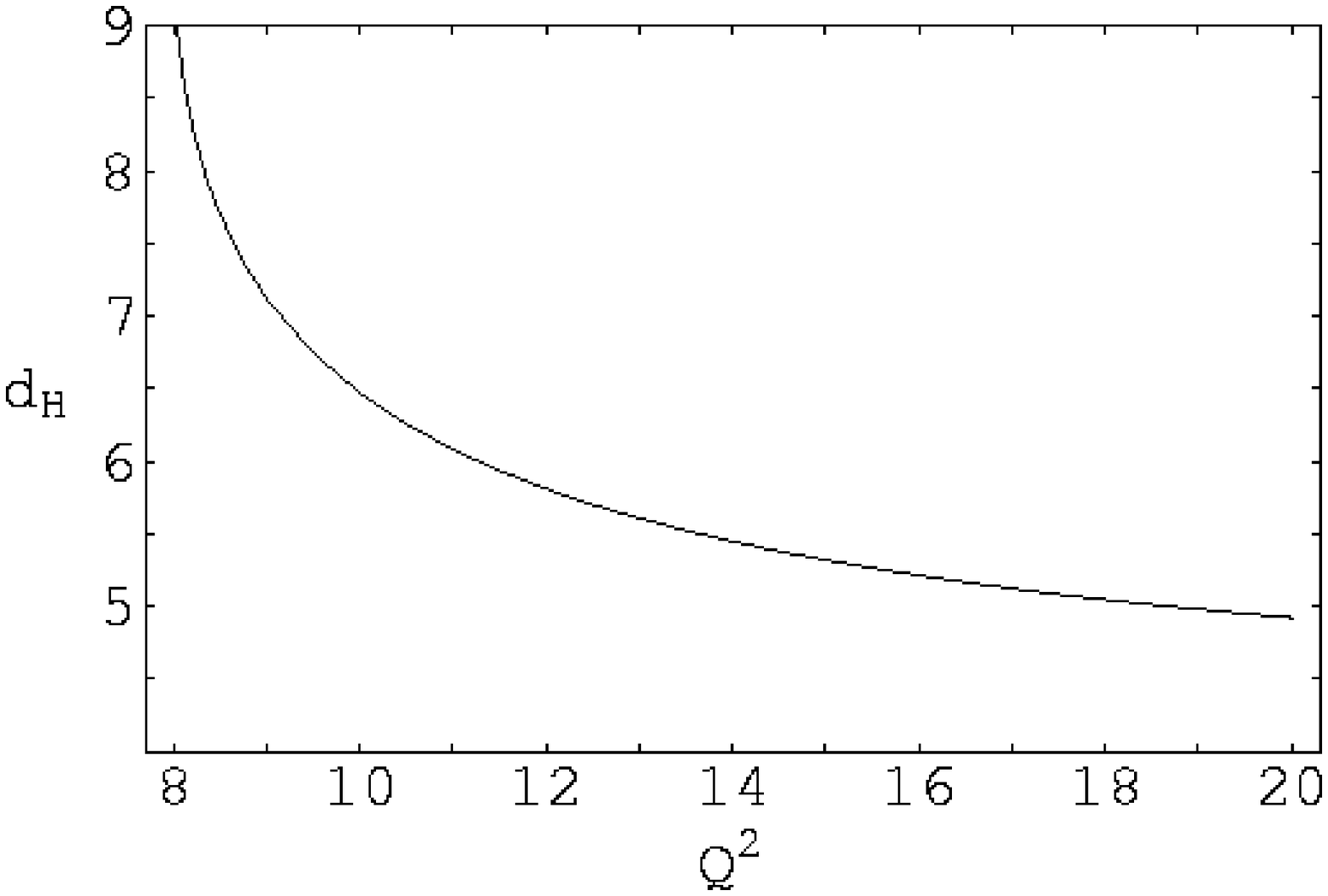}}
\vspace{-1cm}
{FIG. 1. \small{The fractal Hausdorff dimension $d_H$, Eqns. (\ref{hsdf}) and (\ref{dhf}), as a function of the anomaly coefficient $Q^2$.}}

\vspace{3cm}
\epsfxsize=10cm
\epsfysize=9cm
\centerline{\epsfbox{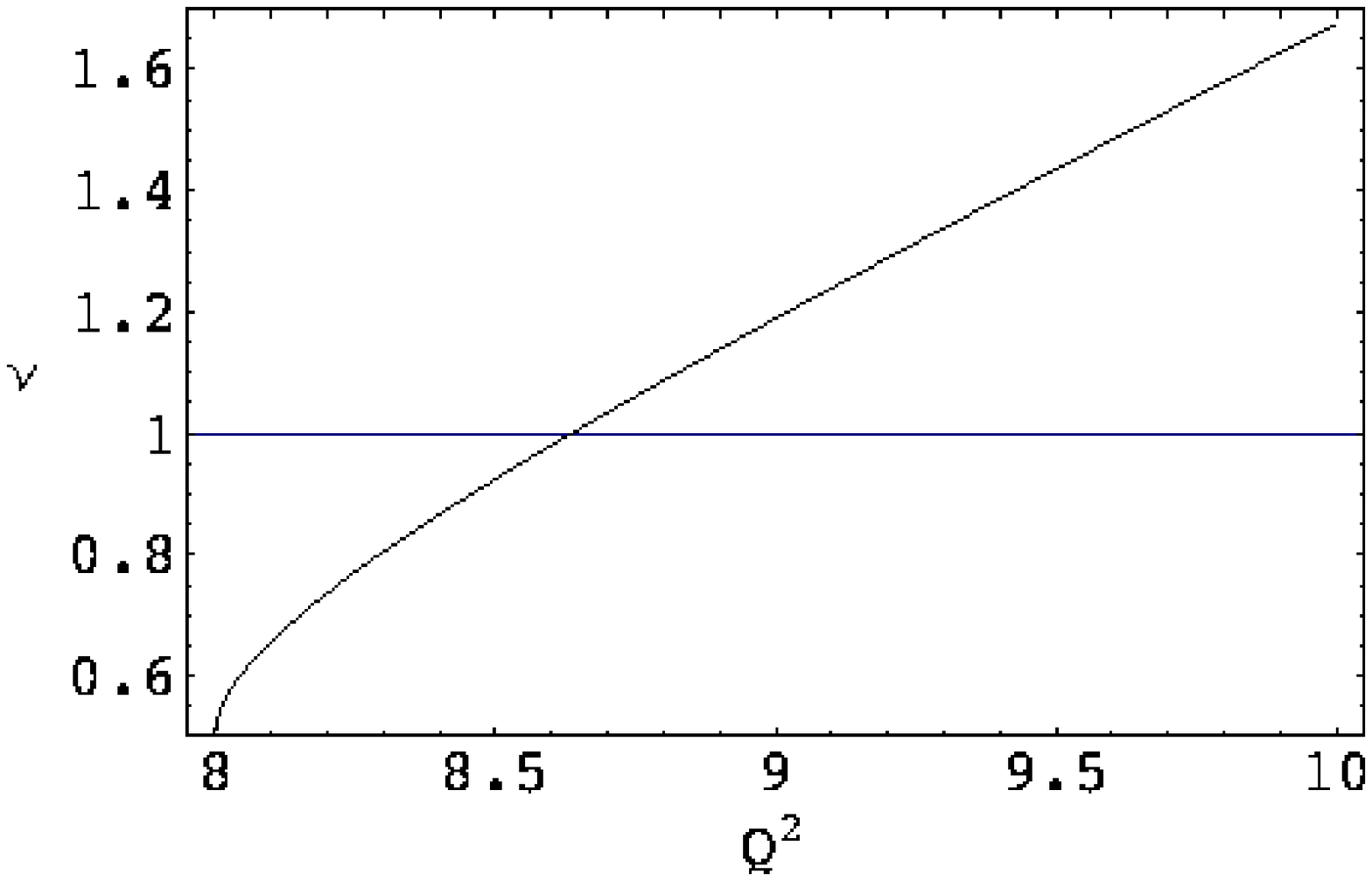}}
\vspace{-2cm}
{FIG. 2. \small{The power law coefficient of the expansion rate of
the universe, Eqns. (\ref{frw}) and (\ref{apower}) as a function of the anomaly coefficient $Q^2$.}} 


\begin{thebibliography}{99}

\bibitem{KPZ} V. G. Knizhnik, A. M. Polyakov, and A. B. Zamolodchikov, 
{\it Mod. Phys. Lett.} {\bf A3} (1988) 819; \hfill\break 
F. David, {\it Mod. Phys. Lett.} {\bf A3} (1988) 1651; \hfill\break
J. Distler and H. Kawai, {\it Nucl. Phys.} {\bf B321} (1989) 509.
\bibitem{DT} F. David, {\it Nucl. Phys.} {\bf B257} (1985) 45; \hfill\break 
V. A. Kazakov, {\it Phys. Lett.} {\bf B150} (1985) 282; \hfill\break 
J. Ambj{\o}rn, B. Durhuus, J. Frohlich and P. Orland, {\it Nucl.
Phys.} {\bf B270} (1986) 457; \hfill\break 
A. Billoire and F. David, {\it Nucl. Phys.} {\bf B275} (1986) 617; \hfill\break 
D. V. Boulatov, V. A. Kazakov, I. K. Kostov and A. A. Migdal, {\it
Nucl. Phys.} {\bf B275} (1986) 641;\hfill\break
J. Ambj{\o}rn and J. Jurkiewicz, {\it Phys. Lett.}
{\bf B278} (1992) 42; \hfill\break 
M.E. Agishtein and A.A. Migdal, {\it Mod. Phys. Lett.} {\bf A7} (1992) 1039;
\hfill\break
S. M. Catterall, J. B. Kogut, and R. L. Renken, {\it Phys. Rev.} {\bf D45}
(1992) 2957.
\bibitem{Rie} R. J. Riegert, {\it Phys. Lett.} {\bf B134} (1984)
56;\hfill\break
E. S. Fradkin and A. A. Tseytlin, {\it Phys. Lett.} {\bf B134} (1984)
187; \hfill\break
I. L. Buchbinder, S. D. Odintsov and I. L. Shapiro, {\it Phys. Lett.} 
{\bf B162} (1985) 93; \hfill\break
E. T. Tomboulis, {\it Nucl. Phys.} {\bf B329} (1990) 410;\hfill\break 
S. D. Odintsov and I. L. Shapiro, {\it Class. Quant. Grav.} {\bf 8} (1991) L57;
\hfill\break S. D. Odintsov, {\it Z. Phys.} {\bf C54} (1992) 531.
\bibitem{am} I. Antoniadis and E. Mottola, {\it Phys. Rev.} {\bf D45}
(1992) 2013.
\bibitem{York} J. W. York, {\it Phys. Rev. Lett.} {\bf 26} (1971) 1656;
{\it ibid.} {\bf 28} (1972) 1082; {\it Jour. Math. Phys.} {\bf 13} (1972) 125;
{\it ibid.} {\bf 14} (1973) 125.
\bibitem{amm95} I. Antoniadis, P. O. Mazur and E. Mottola,
{\it Phys. Rev.} {\bf D55} (1997) 4756; {\it ibid.} 4770.
\bibitem{amm94} I. Antoniadis, P. O. Mazur and E. Mottola,
{\it Phys. Lett.} {\bf B323} (1994) 284; {\bf B394} (1997) 49.
\bibitem{amm97} I. Antoniadis, P. O. Mazur and E. Mottola, 
{\it Phys. Rev. Lett.} {\bf 79} (1997) 14.
\bibitem{Wat} N. Kawamoto, {\it Phys. Rev. Lett.} {\bf 68} (1992) 2113;
\hfill\break 
N. Kawamoto, V. A. Kazakov, Y. Saeki, and Y. Watabiki, {\it Nucl. Phys.}
{\bf B26}({\it Proc. Suppl.}) (1992) 584; \hfill\break 
Y. Watabiki, {\it Prog. Theor. Phys. Suppl.} {\bf 114} (1993) 1;
e-print archive, hep-th/9605185.
\bibitem{Amb} S. Catterall, G. Thorleifsson, M. Bowick, and V. John,
{\it Phys. Lett.} {\bf B354} (1995) 58; \hfill\break 
J. Ambj\o rn and K. N. Anagnostopoulos, {\it Nucl. Phys.} {\bf B497} 
(1997) 445; \hfill\break 
J. Ambj\o rn, K. N. Anagnostopoulos, and G. Thorliefsson, 
e-print archive, hep-lat/9709025.
\bibitem{amm92} I. Antoniadis, P. O. Mazur and E. Mottola, {\it Nucl. Phys.}
{\bf B388} (1992) 627.
\bibitem{DT4} J. Ambj{\o}rn, S. Jain, J. Jurkiewicz and C. F.
Kristjansen, {\it Phys. Lett.} {\bf B305} (1993) 208; \hfill\break
S. M. Catterall, J. B. Kogut, and R. L. Renken, {\it Phys. Lett.} {\bf B328}
(1994) 277; \hfill\break 
B. De Bakker and J. Smit, {\it Nucl. Phys.} {\bf B439} (1995) 239;\hfill\break 
J. Ambj{\o}rn and J. Jurkiewicz, {\it Nucl. Phys.}
{\bf B451} (1995) 643; \hfill\break 
J. Ambj{\o}rn, J. Jurkiewicz, and Y. Watabiki, {\em Jour. Math. Phys.} 
{\bf 36} No. 5 (1995) 6299.
\bibitem{cos} Y. Mellier {\it et. al.}, e-print archive, astro-ph/9609197;
\hfill\break 
A. G. Riess {\it et. al.}, e-print archive, astro-ph/9805201.
\end{thebibliography}
\end{document}